\documentstyle[12pt,openbib,times,psfig]{article}
\newcommand{\al}{\alpha}

\newcommand{\eps}{\epsilon}

\newcommand{\la}{\lambda}

\newcommand{\f}{\frac}

\def\be{\begin{equation}}
\def\ee{\end{equation}}

\begin{document}
\date{}
\title{Bound for entropy \& viscosity ratio of strange quark matter.}

\author{Manjari Bagchi $^{1}$, Jishnu Dey $^{2,\ddagger,
\dagger}$, Mira Dey $^{2,\ddagger,*}$,\\ Taparati Gangopadhyay
$^{2,*}$, Sibasish Laha $^{2,*}$  Subharthi Ray $^{3}$ \& Monika
Sinha $^{4}$}

\maketitle

\noindent {$^1$ Tata Institute of Fundamental Research, Colaba, Mumbai 400005, India \\
$^2$ Dept. of Physics, Presidency College, Kolkata 700073, India\\
$^3$ Inter University Centre for Astronomy \& Astrophysics
(IUCAA), Pune 411007, India \\
$^4$ Theory
division, Saha Institute of Nuclear Physics, 1/AF Bidhannagar, Kolkata 700 064, India \\
$^*$ Supported by DST Ramanna grant, Govt. of India.\\
$^\ddagger$ Associate, IUCAA, Pune, India \\
$^\dagger$ CSIR Emeritus Professor. }

\begin{abstract}

High energy density ($\eps$) and temperature (T) links general
relativity and hydrodynamics leading to a lower bound for the ratio
of shear viscosity ($\eta$) and entropy density ($s$). We get the
interesting result that the bound is saturated in the simple model
for quark matter that we use for strange stars at the surface for
$T~\sim~80~MeV$. At this $T$ we have the possibility of cosmic
separation of phases. At the surface of the star where the pressure is zero - the density $\eps$ has a fixed value for all stars of various masses with correspondingly
varying central energy density $\eps_c$. Inside the star where this
density is higher, the ratio of $\eta/s$ is larger and are like the
known results found for perturbative QCD. This serves as a check of
our calculation. The deconfined quarks at the surface of the strange
star at $T~=~80~MeV$ seem to constitute the most perfect interacting
fluid permitted by nature.

\end{abstract}

\section{Introduction}

~~~~Strange stars are made of deconfined $u,~d,~s$ matter. The
pressure at the star surface is zero with a surface number density is around $4-5$ times
the normal matter density. The central density is almost 3 times
the surface density. We find that the ratio of the kinetic viscosity
to entropy density of strange stars (SS) nearly saturates the lowest
possible bound found by Kovtun, Son and Starinets \cite{kss} (KSS in
short) at the surface at high T. This is as perfect as an
interacting fluid can be. The relevant T where this happens is where
the cosmic separation of phases takes place \cite{wit}. This is in
the sense that it is the critical T above which no zero pressure
point exist for the deconfined quarks. This implies  that above this T there can be no self bound strange stars. Below this T, the two phases of hadronic stars and quark stars can both exist \cite{brdd} as the surface tension of the strange stars is high
\cite{AALett}. The temperature estimated by Witten \cite{wit} for
this was $T~=~100~MeV$ which is close to what we get.

Our calculation is surprising to a certain extent, we try to confirm
it by moving from the surface to the inside of the strange star. KSS
state that somewhat counterintuitively, a near ideal gas has a large
viscosity. In agreement with this observation, deep inside the star
the condition are more like perturbative (or weak coupling) QCD and
we find that $\eta/s$ is larger than at the surface and comparable
to the results of Arnold, Moore and Yaffe \cite{amy}. This is a
consequence of the crucial density dependence of the quark mass that
we have assumed and can be interpreted as a support of our
assumption. We must stress however that the value of strong coupling
constant $\al_s$ relevant for the KSS bound is large $\sim 0.6$.

We talk of shear viscosity that is relevant for the problem and the
bulk viscosity is negligible at least for weak coupling as shown by
Arnold, Dogan and Moore \cite{adm}. For values of $\al_s~ \sim~0.3$
the bulk viscosity is thousand times smaller that the shear
viscosity. Interestingly they note that at high density where the
QCD coupling is small, there are long lived quasiparticles and a
kinetic theory treatment should be valid which we find to be valid
also at larger $\al_s$.

Many of the relevant points discussed in the literature are
summarized in a recent review by Blaizot \cite{bla}. The
experimental data from heavy ion collisions (RHIC) do not provide
any evidence for ideal gas behavior,  rather the produced matter
behaves as a fluid with low viscosity, the ``perfect fluid".

New techniques have emerged that allows calculations to be done in
some strongly coupled gauge theories that differs however in
essential aspects from QCD. The answer to the question - is quark
-gluon plasma weakly or strongly coupled -  does not have a straight
forward answer. Indeed in the quark gluon plasma coexist seemingly
perturbative features, and non perturbative ones. This is the view
which matches with our spirit.

The background for the viscosity bound conjecture of KSS \cite{kss}
will be briefly touched upon for the sake of completeness :

It is popularly known that black holes are endowed with
thermodynamics. In higher dimensional gravity theories there
exist solutions called black branes and they are black holes with
translationally invariant horizons. For these solutions
thermodynamics can be extended to hydrodynamics - the theory that
describes long-wavelength deviations from thermal equilibrium.
Applying the holographic principle a black brane corresponds to a
certain finite-temperature quantum field theory in fewer number
of space time dimensions, and the hydrodynamic behaviour of
black-brane horizon is identical with the hydrodynamic behaviour
in a dual theory.

The arguments of KSS for  generalization of the viscous bound
$4~\pi~\eta/s~>~ 1$ - is more interesting since it only invokes
general principles like the Heisenberg uncertainty relation for the
typical mean free time of a quasi-particle and the entropy density
$s$. From here to our model is just one short step of identifying
the quasi-particles to be the dressed quarks of a mean field
description for a large colour effective theory. Further light in
this direction comes from the recent work of Fouxon, Betschart and
Bekenstein (FBB in short) \cite{fbb} as we shall discuss later in
this paper. For a black hole calculation for the matter inside is of
course impossible so FBB concentrate on the generalized second law
of thermodynamics that they call GSL. Following them one can state
that GSL claims that the sum of entropy of all the black holes and
the total ordinary entropy in the black holes' exterior never
decreases. Then they go on to consider a simple spherical accretion
model and suggests that this Bondi flow satisfies GSL  because the
accretion velocity approaches the speed of light.

Our model is presented in the next section emphasizing the possible
astrophysical observational checks that have already been discussed
extensively in the literature. In section 3, we describe the
calculation of the viscosity known to all. In section 4, the
considerations enumerated by FBB are shown to be satisfied in our
model and we present a summary and conclusion in the last section.

\section{Strange stars at finite T}

~~~~The density dependent quark mass is given in our model as :

\be M_i = m_i + M_Q ~sech\left(\frac{n_B}{N n_0}\right), \;\;~~~ i =
u, d, s. \label{eq:qm} \ee
 where $n_B = (n _u+n _d+n _s)/3$ is the
baryon number density, $n_0 = 0.17~fm^{-3}$ is the normal nuclear
matter density, and $N$ is a parameter taken to be 3 in the set F of
\cite{brdd} which we have chosen here. The results for A-E are not
too different as can be seen from Table 1 of \cite{brdd}. For set F
the maximum mass possible for SS is $1.436 M_\odot$ and the corresponding
radius is $6.974\, km$. At high $n_B$ the quark mass $M_i$ falls from a
large value $M_Q$ to its current one $m_i$ which we take to be
$m_u = 4 \;MeV,\; m_d = 7 \;MeV,\; m_s = 150 \;MeV$  \cite{D98}.
$M_Q$ is taken as $345 ~MeV$ in set F of \cite{brdd}. Possible variations of chiral symmetry restoration at high density (CSR) can be incorporated in the model
through $N$.

We use a modified Richardson potential with different
scales for confinement ( $\sim ~350~MeV$ ) and asymptotic freedom
($100~MeV$) which has been tested by fitting the octet and decuplet masses and
magnetic moments \cite{man1, man2} and the temperature dependence of
the gluon mass is taken from Alexanian and Nair \cite{an}.

The finite $T$ calculation involves a $T$-dependent gluon
screening and thermal single particle Fermi functions with
interactions that involve all pairs of quarks. Along with the
painstaking constraints of $\beta $ - equilibrium and charge
neutrality in these calculations - it is found  that zero pressure occurs at a density $\sim$ 4 to 5 times the normal nuclear
density $n_0$ till $T~=~80~MeV$. This is a relativistic mean field
calculation with a screened Richardson potential for two quarks,
where only the Fock term contributes. The calculation is self
consistent. Strange quark matter is self bound by strong
interaction itself. The energy density and pressure of this
matter lead to strange quark star through the TOV equation with
mass and radius depending on the central density of the star.

The model has been applied to discussions on compactness of stars
\cite{D98, prl, apjl}, quasi-periodic oscillations in X-ray power
spectrum \cite{bani}, the existence of minimum magnetic field for
all observed pulsars \cite{raka}, absorption and emission bands
along with high redshift \cite{manjari1}, superbursts \cite{monika}
and high value of surface tension useful to stabilize the strange
stars \cite{AALett}.

\section{Calculations}

~~~~We use the classical expression for evaluating the shear
viscosity coefficient $\eta$ as:

\be \eta=\frac{1}{3} mv n \lambda \label{kin} \ee
 where the mean free path $\lambda$ is given in terms of the interaction
diameter of quark $d_q$ and the appropriate number density n

\be
 \lambda = \f{1}{(4/3) n d_q^2 }. \label{freepath}
\ee

We need to specify the average momentum $P$ which we take from the
Fermi distribution

\begin{equation}
\langle P\rangle = mv = \f{\int_0^\infty k^3 f(k, U_i)dk}{\int_0^\infty k^2 f(k, U_i)dk},~~~~ i = u,d,s
\label{mom}
\end{equation}

\begin{equation}
 f(k,U_i)= \f{1}{1~+~exp[(U_i - \mu_i)/T]}.
\label{fermi}
\end{equation}

\begin{figure}[htbp]
\centerline{\psfig{figure=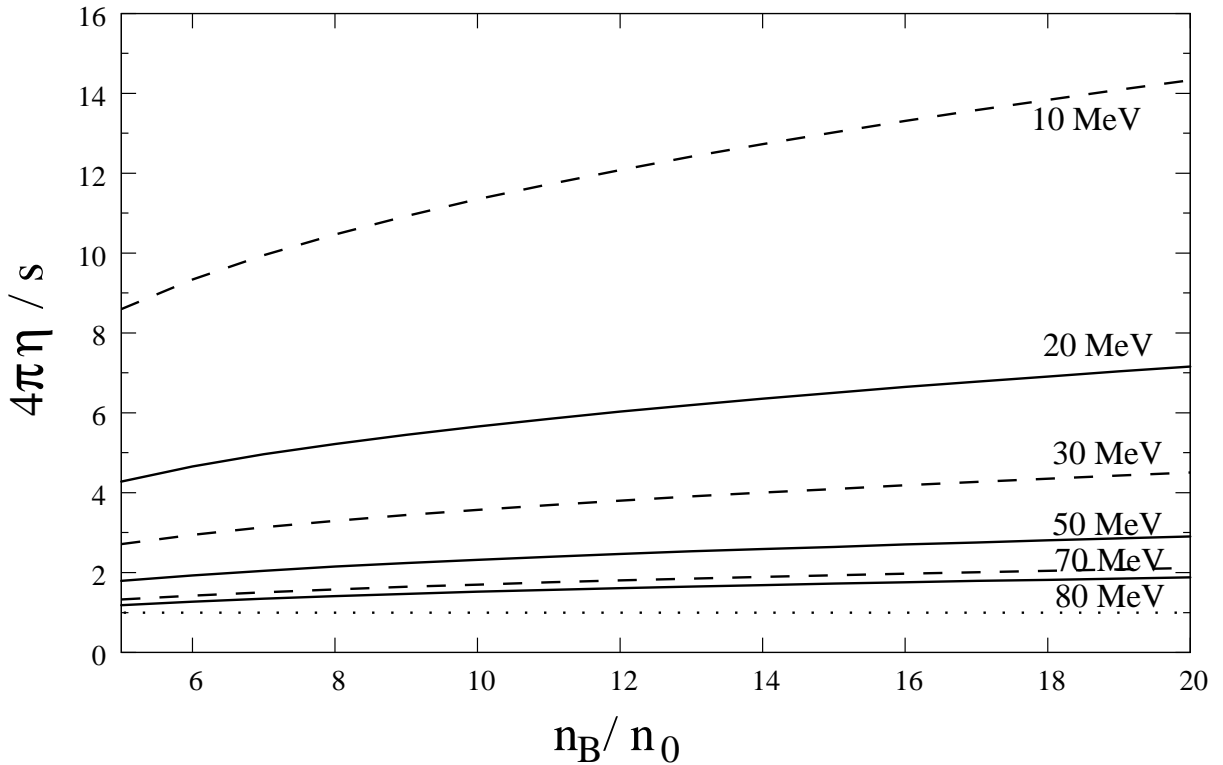}} \caption{$4\pi$ times shear
viscosity $\eta$ divided by the entropy density for various number
density is plotted. According to the KSS bound \cite{kss} this
should be one for what is called the most perfect fluid, perhaps
encountered in RHIC \cite{lacey}. We see that the bound is nearly
saturated at $n_B/n_0\sim 5$ which is the surface of the star at T =
$80~MeV.$} \label{fig:etasvsnum}
\end{figure}

Heiselberg and Pethick(1993) suggested that the quark scattering
cross section $\pi r^2$ can be compared to proton-proton scattering
using the quark counting rule
$\sigma_{pp}$~=~$3~\sigma_{qq}$~=~$3\pi r_{n}^2$ \cite{Heisel} where
$r_n$ is the interaction radius. In matter this is calculated by
assuming that the relevant particles (in this case the quarks)
occupy an effective volume $\frac{4}{3}\pi~r_n^3$.

We calculate the diameter of the quarks $d_q$ by assuming that they
are packed tightly on the surface of the star. This is justified
since the gravitation is strong and it will try to minimize the
surface. The quarks, assumed to be spheres, have radius
$r_q~=~d_q/2$ and their projected area on the surface of the star
($4\pi~R^2$) will be $\pi~r_n^2$ giving the number to be :

\be N_q = 4 R^2/r_n^2 \label{Ns}. \ee The volume of the tightly
packed layer is $V~ =~ 4\pi~R^2 \times d_n$ and the number is $V
\times n$ where the $n$ is the self consistent number density
corresponding to the definition of the zero pressure surface of our
model. This number, equated to $N_q$ given above, leads to : \be
 d_n = \left[\frac{4}{\pi n}\right]^{1/3},
\label{diameter}
\ee

The number density for the strange star in our model changes from
the surface where it is between four and five times the normal
nuclear matter density $n_0$ to about 15 times $n_0$ in the centre
of the star for T = 0. For finite T the numbers increase somewhat
due to the Fermi distribution.

We see in Fig.(\ref{fig:etasvsnum}) that the $4\pi\eta/s \sim 1$ for
the highest T where strange stars are self bound for the star
surface which has the lowest value of the number density. At higher
densities the ratio is much larger as is the case for perturbative
QCD.

The variation of $\eta/s$ with the coupling is counter-intuitive as
emphasized by KSS. We wanted to check that the ratio in fact
increases with decreasing coupling. To do this we needed the
relevant $\al_s$ at each density.

We have extracted the strong coupling constant $\al_s$ from the
density dependence of the mass given in eq.(\ref{eq:qm}) as in
\cite{ray,rdd}. This is due to the simplified Schwinger-Dyson
formalism of Bailin, Cleymans and Scadron using the Dolan-Jackiw
Real Time propagator for the quark. We re-do the calculation here for
the $M_d$ and the $n~=~3$ appropriate for our latest parameter set
but essentially there is no fundamental change in $\al_s$, the
variation being from $\sim 0.6$ at low density to about 0.2 at the
highest.

\be \al_s(r,n) = \f{m_{dyn}-M_d(r,n)\pi}{2~
m_{dyn}~\ln[\f{\mu(r,n)+(\mu(r,n)^2-M_d(r,n)^2)^{.5}}{M_d(r,n)}]}.
\label{dyson} \ee\

\begin{figure}[htbp]
\centerline{\psfig{figure=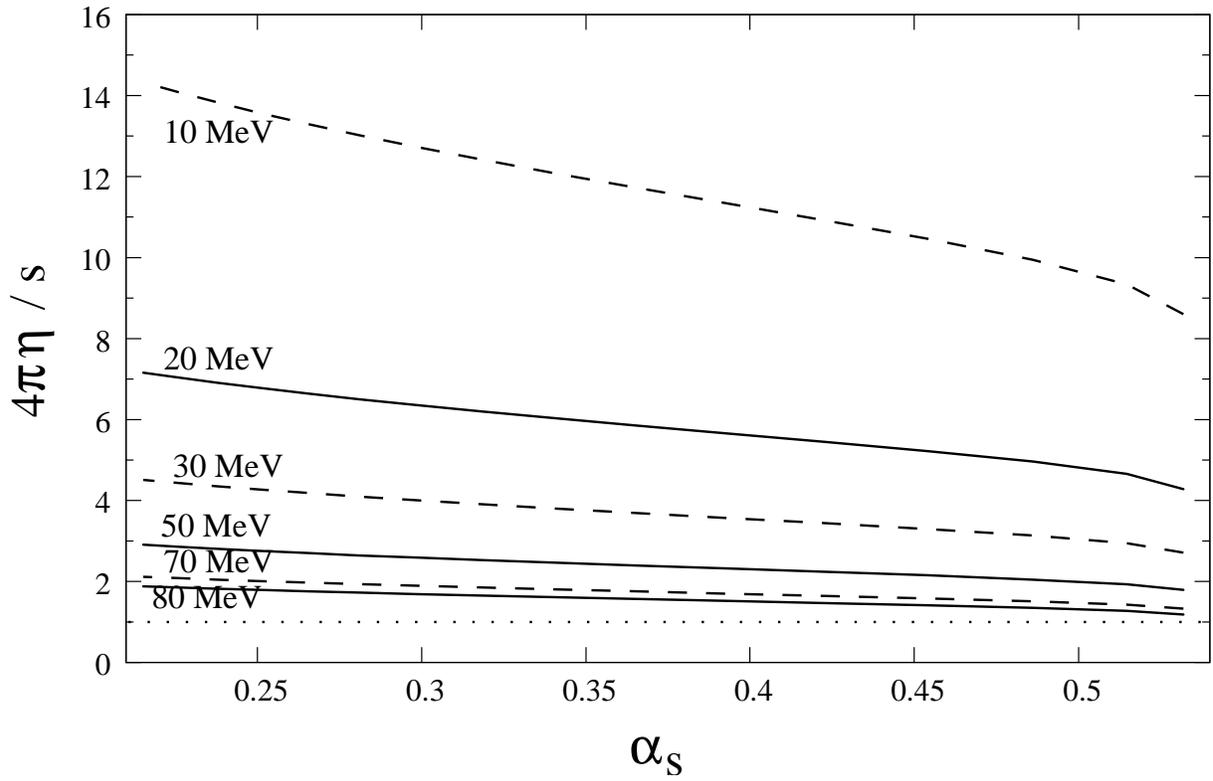}} \caption{We also find that
$\eta$ is a decreasing function of coupling strength as discussed
for example  by Stephanov \cite{stephanov}. We should stress that
the value of $\al_s$ relevant for this paper is large, about 0.65.}
\label{fig:etasvsalpha}
\end{figure}

The variation of $4\pi\eta/s$ with $\alpha_s$ has been shown in fig.
\ref{fig:etasvsalpha}. The interesting point here is that the value
of $4\pi\eta/s$ is larger than one by factors ranging from 2 to 14
for various T at $\alpha_s~\sim~0.2$ so that it is clear that
transport of quarks is the main factor for the largeness of this
factor and the smallness of the interaction does not matter.

In a recent paper Lacey has given a very lucid and colourful
representation of viscosity bound for different fluids (see fig. 3
of \cite{lacey}) which we summarize here. As the $(T-T_c)/T_c$
varies from ~-0.5~ to 0, $\eta/s$ in (a) meson gas goes from 1.2 to
0.4, (b) water goes from 3.8 to 2.2 (c) liquid nitrogen from  3.4 to
0.8 and (d) liquid helium from 3.4 to 0.8. The matter in the strange
star seems to be the first so called perfect interacting liquid
where bound reaches the fraction $\sim (4\pi)^{-1}$ and thus it may
be the same fluid which Lacey marks as RHIC.

We would like to mention another recent paper dealing with
boost-invariant viscous hydrodynamics \cite{janik} for although this
deals with a theory which does not have a direct counterpart which
works for QCD it may still be useful for studying features of the
plasma that is strongly coupled and deconfined.

\section{Bekenstein bound \& its connection to that of KSS.}

~~~~In a recent paper FBB \cite{fbb} has suggested that the KSS
bound is related to the Bekenstein bound \cite{bek}

\be S/E < 2\pi R\ee where $R$ is the radius of the smallest sphere
circumscribing a system whose entropy is $S$ and energy is $E$ and
then they reduce it to what they call the UBE, the universal bound
for entropy :

\be
s/\eps < 2\pi \lambda
\label{UBE}
\ee
 where $s, \eps$ are the entropy and energy densities respectively
and $\lambda$ is the mean free path. In Table(\ref{tab:bound}) we
present these quantities and it is clear that the inequality is
satisfied T increases from 1 to 80 $MeV$, at the surface where quark
number density varies from $2.04~fm^{-3}$ to $3.22~fm^{-3}$. At T = 90
$MeV$, which is the last entry, the eqn. (10) is just about violated
and coincidentally a zero pressure point is no longer there in our
equation of state.

The use of Bekenstein bound for RHIC is not new. The entropy bound
has been invoked to set limits for T at which hadrons can survive as
a confined system. For example, the pion may form at lower T than
the $\rho$ meson \cite{ddst} and that the pion cannot exist at 90
$MeV$ if its mass is 138 $MeV$ and and its radius is 0.445 $fm$ (see
Table 2 of \cite{ddst}). It is satisfying to see that the same
temperature is invoked in strange quark matter with the updated
Bekenstein bound Table (\ref{tab:bound}).

\begin{table}[h]
\caption{Comparing entropy-energy ratio with momentum at different
temperature T. It may be noted that the Bekenstein bound as updated
by FBB, namely $s/\eps \le 2\pi \lambda$ is exactly satisfied as an
equality between T = 80 and 90 (in $MeV$). Number density is $n$,
$d_q$ is the average interaction diameter of the quarks at the star
surface, $P$ is the average momentum and $\eta$ is the kinetic
viscosity. }
\begin{center}
\begin{tabular}{|c|c|c|c|c|c|c|c|c|c|}
\hline T& $n(fm^{-3})$ & $\eps(fm^{-4})$ & $s(fm^{-3})$ & $d_q(fm)$ & $\eta(fm^{-3})$ & $P(fm^{-1})$
 & $2\pi\la$ & $s/\eps$ & $\eta/\eps\la$ \\
\hline

1  & 2.04   & 3.202  & .05615 & .85459 & .45972 & 4.2192 & 1.0068 & .01754 & .89599 \\
10 & 2.0549 & 3.3528 & .57913 & .85251 & .46544 & 4.2509 & 1.0043 & .17273 & .86845 \\
20 & 2.1023 & 3.5843 & 1.1927 & .84606 & .48454 & 4.3585 & .99674 & .33276 & .85215 \\
30 & 2.1846 & 3.9036 & 1.8386 & .83530 & .51768 & 4.5390 & .98407 & .47101 & .84672 \\
40 & 2.3019 & 4.3221 & 2.5221 & .82087 & .56420 & 4.7774 & .96706 & .58351 & .84812 \\
50 & 2.451  & 4.8384 & 3.2535 & .80387 & .62202 & 5.0511 & .94704 & .67242 & .85292 \\
60 & 2.6255 & 5.438  & 4.0445 & .78566 & .68798 & 5.3365 & .92558 & .74375 & .85882 \\
70 & 2.8174 & 6.0988 & 4.9062 & .76739 & .75872 & 5.6147 & .90406 & .80446 & .86459 \\
80 & 3.0193 & 6.7975 & 5.8484 & .74989 & .83129 & 5.8743 & .88344 & .86037 & .86976 \\
90 & 3.2249 & 7.5131 & 6.8796 & .73360 & .90340 & 6.1096 & .86425 & .91568 & .87417 \\

\hline
\end{tabular}
\end{center}
\label{tab:bound}
\end{table}

One can proceed to find more interesting results. According eqn.
(34) of FBB, $\eta \sim \eps \la a$ where $a$ is the speed of sound.
Thus the last column of Table (\ref{tab:bound}) shows that the
velocity of sound is close to the velocity of light. This is
consistent with the findings of Sinha $et~al$ \cite{monika2} where
$a$ is calculated from first principles by evaluating the
incompressibility. As stated in our introduction luminal velocity of
Bondi accretion flow  $U_{ac}~\sim~1$ was invoked by FBB and this is
reminiscent of that.

At T = 80 $MeV$ we have \be s~=~4\pi~\eta~=~(4\pi/3)P~n\lambda<2\pi
\eps \lambda \ee which yields the inequality for the average
momentum $P~<~1.5 \eps/n$ where $P$ is the average momentum and
$\eps/n$ is the energy per particle. This can be directly
compared with KSS who state that the energy of a quasiparticle and
its mean free time $\tau_{mft}$ cannot be smaller that $\hbar$ and
hence $\eta/s~\ge~\hbar/k_B~$. Recalling that we work with units
$k_B~=~\hbar~=~c~=~1$ and that the quarks have velocities comparable
with the velocity of light $c$ one can see that both relations are
consistent with the uncertainty relation. Thus it can be asserted
that the generalized second law of thermodynamics and the
uncertainty relation have some consistency checks if one uses the
Bekenstein bound UBE and the KSS bound.

\section{Discussion}

We are grateful to the anonymous referee for raising an
important question that what happens at a temperature higher than  $\sim ~100~ MeV$ or a density much lower than 4-5  times the normal matter
density ? The deconfined strange quark matter does not exist below the critical density of  4-5  times the normal matter density above a temperature of 80 MeV in our mean field model. In Witten's original scenario \cite{wit}  for cosmic separation of phases - a QCD and a hadron phase started around 100 MeV. A different phase was
implied above this temperature which was not specified. One could imagine this
could be a pre-QCD phase or it could be hadrons overlapping  with
quarks percolating through. We propose that the
hydrodynamics  of such  a  phase  will satisfy the KSS bound
along the boundary of the density-temperature curve on which our
point is a low temperature high density point whereas in RHIC a
lower density and a higher temperature of 200 MeV may be obtained
and will show the KSS bound. It is our conjecture that the KSS
bound is always valid on this curve. To us this seems to be a
likely scenario in view of the many model calculations done by
many groups recently \cite{chen, csernai}.

\section{Summary and Conclusions}

$\eta$ increases with increasing energy density $i.e.$ decreasing
$\al_s$ for the matter that composes a self bound strange star. The
transport here is radial hence $\eta$ is the shear viscosity. At the
surface of the star the pressure is zero and  the number density is
the same for stars of all masses. The quark matter at the surface
saturates the bound given by \cite{kss} for $T~=~80~MeV$ - the
highest T where we get zero pressure.

Our model leads to such an interesting result, connecting zero
pressure with the viscosity bound on the one hand and RHIC on the
other hand. The updated Bekenstein bound is exactly satisfied as an
equality at high density between T = 80 and 90 $MeV$ where Witten's
cosmic separation of phases is possible.
%%%%%%%%%%%%%%%%%%%%%%
\section*{Acknowledgments}

The authors TG, MB, MD and JD are grateful to IUCAA, Pune, and
HRI, Allahabad, India, for short visits. We are grateful to
Rajesh Gopakumar for drawing our attention to the paper by Kovtun,
Son and Starinets.

%%%%%%%%%%%%%%%%%%%%%%%%%%%%%%%%%%%%%%%%%%%%%%%%%%%%%%%%%%%%%%%%%%%%%%%%%
%%%                REFERENCES
%%%%%%%%%%%%%%%%%%%%%%%%%%%%%%%%%%%%%%%%%%%%%%%%%%%%%%%%%%%%%%%%%%%%%%%%%

\end{document}